\documentclass[a4paper]{article}
\usepackage[pdftex]{graphicx} 
\usepackage[english]{babel}
\usepackage[latin1]{inputenc} 
\usepackage[bf,small,nooneline]{caption}

\title{Modeling the Lunar plasma wake}
\date{June 29, 2012}
\author{M.\ Holmstr{\"o}m\thanks{Swedish Institute of Space Physics, PO~Box~812, SE-98128~Kiruna, Sweden. (\texttt{matsh@irf.se})},  B.\ Eliasson\thanks{Ruhr-University Bochum, Faculty of Physics and Astronomy, Bochum, Germany.}}

\begin{document}
\maketitle

%
%
%

\section{Introduction}
Solar system bodies that lack a significant atmosphere and internal 
magnetic fields, such as the Moon and asteroids, can to a first 
approximation be considered passive absorbers of the solar wind. 
The solar wind ions and electrons directly impact the surface of these bodies,  
and a wake is created behind the object. 
In a frame that moves with the solar wind plasma, the refill of 
the lunar wake can be viewed as the more general problem of plasma 
expansion into a vacuum. 

One-dimensional plasma expansion into a vacuum has been studied 
for a long time, see~\cite{Mora} and references therein. 
The specific case of one-dimensional plasma expansion into the lunar 
wake has also been studied~\cite{Farrell98,Birch01}. 
A disadvantage of one-dimensional models is that the magnetic 
field component along the dimension is forced to be constant, from the 
requirement that $\nabla\cdot\mathbf{B}=0$. 
On the other hand, three dimensional models~\cite{eps} does not 
allow the same spatial resolution and size of the simulation domain 
as lower dimensional models, due to computational constraints. 
Thus, a two-dimensional model of the lunar wake refill could be useful. 

A one-dimensional approximation of the lunar wake is an initial 
state with uniform plasma, corresponding to the solar wind, 
surrounding a region of vacuum (zero density) of width equal to 
the lunar radius.  With time the plasma will expand into the 
vacuum region.  This approximates plasma conditions along a line 
that at $t=0$ lies in the terminator plane and goes through the 
center of the Moon.  The line then convects with the solar wind 
as time goes by. 
The magnetic field at $t=0$ corresponds to the 
interplanetary magnetic field (IMF) in the solar wind. 
It is possible to use the same convecting frame approach 
also in two dimensions. 
If we consider the terminator plane at $t=0$ and let this plane 
convect with the solar wind plasma bulk velocity, $u_{sw}$, it will 
at time $t$ be at $x=-t u_{sw}$.  The initial configuration at $t=0$ is the 
$yz$-plane with proton number density $n=n_0$ for $r<R$, 
and $n=n_{sw}$ for $r>R$. 
Here the $x$-axis is anti-parallel to the solar wind, and $r=\sqrt{y^2+z^2}$. 
We study such plasma expansion into a vacuum in a two-dimensional 
geometry using a hybrid model with $R=1730$~km. 
The hybrid model has ions as particles and electrons as a mass-less 
fluid, and is described in~\cite{eps}, and references therein. 
In the hybrid model the electric field computation involves a division by 
charge density.  Thus, there is a problem in vacuum regions. 
Here we handle this by letting $n_0$ have a small, non-zero value. 
In this case $n_0=0.1n_{sw}$.  However, this low density population 
of ions is not shown in the plots. 

We can note that more complicated low dimensional models are possible. 
We could follow the convecting plane (or line in the one-dimensional case) 
from upstream, in the undisturbed solar wind, and after each time step 
remove ions that collide with the Lunar surface from the ion velocity 
space distribution.  This removal would continue until the plane has 
convected past the Moon.  The solution would however no more be 
independent on the solar wind velocity.

\section{Results}
In what follows, 
the solar wind number density, $n_{sw}=5$~cm$^{-3}$, 
the ion temperature is $5\cdot10^4$~K, 
the electron temperature is $10\cdot10^4$~K, and 
the IMF is in the $xy$-plane with a magnitude of 5~nT. 
For these plasma parameters, the ion inertial length is 100~km, 
the Alfv\'{e}n velocity is 49~km/s, 
the thermal proton gyro radius is 60~km (the gyro time is 13~s), 
and the ion plasma beta is 0.35. 
The cell size is 78~km and the number of meta particles per cell is 170. 

\begin{figure}
\begin{center}
\includegraphics[width=118mm]{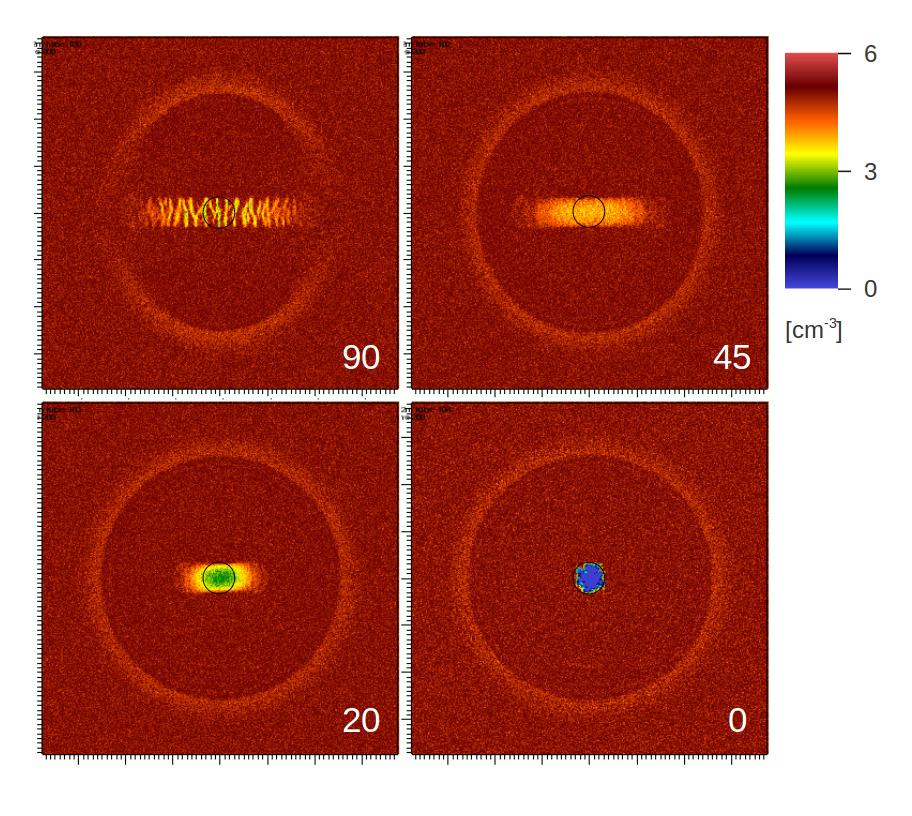}
\end{center}
\caption{Proton number density in the $yz$-plane 
         at $t=200$~s for an IMF direction of 
         90, 45, 20, and 0 degrees away from the image normal 
         (solar wind direction). The plane of the IMF is horizontal. 
         The black circle show the outline 
         of the initial low density region of radius 1730~km 
         (representing the Moon). 
}
\label{fig:imf}
\end{figure}

How the wake number density depends on the direction of the 
initial magnetic field is shown in Fig.~\ref{fig:imf}. 
The time $t=200$~s of the plot corresponds to a distance of $t u_{sw}$
downstream of the moon, e.g., for $u_{sw}=500$~km/s the distance is 
10,000 km $\approx$ 5.8 Lunar radii. 

We see a density rarefaction that expands from the Moon at the 
fast magnetosonic speed, as has been seen before in three-dimensional 
simulations~\cite{eps}. 
It does not propagate parallel to the IMF as can be seen for the 
case of an IMF at a 90~deg angle to the solar wind flow direction. 
At this time, for these condition, 
the wake has partially refilled except for the case of 
a 0~deg IMF (parallel to the solar wind flow) for which we still 
have a (deformed) vacuum region. 
We also see that the wake is flattened in the plane of the IMF due 
to the fact that the ions flow along the IMF. 

For the 90~deg IMF case we see density oscillations in the refilled 
region, where surrounding plasma intrudes. 
In the next section we derive criteria for an instability in the 
hybrid model that can be a possible cause for these oscillations. 
Note that these oscillations are a feature of the hybrid model, 
it is unknown if such oscillations also would occur in a 2D 
particle in cell model of the Lunar wake that includes also 
electrons as particle. 

\section{Ion beam instability in a hybrid model}
For longitudal waves propagating along the magnetic field lines, the hybrid model reads
\begin{equation}
  \frac{\partial f_i}{\partial t}+ v\frac{\partial f_i}{\partial x}+\frac{eE}{m_i}\frac{\partial f_i}{\partial v}=0
\end{equation}
where the electric field is obtained from the electron momentum equation as
\begin{equation}
  E=\frac{k_B T_e}{e}\frac{\partial}{\partial x}\bigg(\frac{n}{n_0}\bigg)^{5/3},
\end{equation}
and quasi-neutrality gives
\begin{equation}
  n=\int f_i d^3v
\end{equation}
Linearization $f_i=f_0+f_1$, $E=E_1$, $n=n_0+n_1$ gives
\begin{equation}
  \frac{\partial f_1}{\partial t}+ v\frac{\partial f_1}{\partial x}+\frac{eE_1}{m_i}\frac{\partial f_0}{\partial v}=0
\end{equation}
\begin{equation}
  E_1=\frac{5}{3}\frac{k_B T_e}{e n_0}\frac{\partial n_1}{\partial x},
\end{equation}
and
\begin{equation}
  n_1=\int f_1 d^3v
\end{equation}
Laplace transformation $\partial/\partial t\rightarrow -i\omega$ and $\partial /\partial x\rightarrow ik$ gives
\begin{equation}
  (-i\omega+ivk)f_1 + \frac{e E_1}{m_i}\frac{\partial f_0}{\partial v}=0,
\end{equation}
and 
\begin{equation}
 E_1=\frac{5}{3}\frac{k_B T_e}{e n_0}i k n_1=\frac{5}{3}\frac{k_B T_e}{e n_0}i k\int f_1 d^3v
 = - \frac{5}{3}\frac{k_B T_e}{m_i n_0}i k E_1 \int \frac{1}{-i \omega+ivk} \frac{\partial f_0}{\partial v} d^3v.
\end{equation}
Eliminating the common factor $E_1$ and rearranging the terms, we obtain the dispersion relation
\begin{equation}
  1+ \frac{5}{3}\frac{k_B T_e}{m_i n_0} \int \frac{1}{v-\omega/k} \frac{\partial f_0}{\partial v} d^3v=0,
\end{equation}
Here the frequency and wavenumber appear only as a ratio $\omega/k$. If we replace this ratio by $v_p$, the "phase velocity", we
obtain the equation
\begin{equation}
  1+ \frac{5}{3}\frac{k_B T_e}{m_i n_0} \int \frac{1}{v-v_p} \frac{\partial f_0}{\partial v} d^3v=0,
\end{equation}
Note that in this equation, $v_p$ depends only on the plasma parameters but not on $\omega$ or $k$. 
Hence $v_p$ is a complex number, $v_p=v_R+iv_I$, which can either have positive or negative imaginary parts $v_I$.
The dispersion relation can be written $\omega=(v_R+iv_I)k$. If $v_I$ is positive, then we have an instability
which formally goes to infinity as $k\rightarrow \infty$. In a simulation, however, the wavenumber is limited by
the grid size. 

\section{Acknowledgments}
This work was supported by the Swedish National Space Board, and 
the Swedish National Infrastructure for Computing. 
The software used in this work was in part developed by the DOE NNSA-ASC 
OASCR Flash Center at the University of Chicago.

\end{document}